\documentclass[11pt]{article}
\usepackage{amsmath}
\usepackage{amsbsy}
\usepackage{eurosym}

\input{tcilatex}

\begin{document}

\begin{center}
\textbf{\large Of the Exterior Calculus and Relativistic Quantum Mechanics }

Jose G. Vargas

January 20, 2019

\bigskip

\textit{To Mr. Eric Alterman, for his funding of events promoting the K\"{a}%
hler calculus}\bigskip
\end{center}

\textbf{Abstract.} In 1960-62, E. K\"{a}hler developed what looks as a
generalization of the exterior calculus, which he based on Clifford rather
than exterior algebra \cite{R46}, \cite{R47} and \cite{R48}. The role of\
the exterior derivative, $du$, was taken by the more comprehensive
derivative $\partial u$ ($\equiv dx^{\mu }\vee d_{\mu }u$), where ``$\vee $%
'' stands for Clifford product. The $d_{\mu }u$ represents a set of
quantities to which he referred as covariant derivative, and for which he
gave a long, ad hoc expression. We provide the geometric foundation for this
derivative, based on Cartan's treatment of the structure of a Riemannian
differentiable manifold without resort to the concept of the so called
affine connections.

Buried at advanced points in his presentations \cite{R46}, \cite{R48} is the
implied statement that $\partial u=du+\ast ^{-1}d$ $u\ast $, the sign at the
front of the coderivative term is a matter of whether we include the unit
imaginary or not in the definition of Hodge dual, $\ast $. We extract and
put together the pieces of theory that go into his derivation of that
statement, which seems to have gone unnoticed in spite of its relevance for
a quick understanding of what his ``K\"{a}hler calculus''.

K\"{a}hler produced a most transparent, compelling and clear formulation of
relativistic quantum mechanics (RQM) as a virtual concomitant of his
calculus. We shall enumerate several of its notable features, which he
failed to emphasize. The exterior calculus in K\"{a}hler format thus reveals
itself as the computational tool for RQM, making the Dirac calculus
unnecessary and its difficulties spurious.

\section{Introduction}

In the preface of his authoritative book ``The Dirac Equation", B. Thaller
states:

\begin{quotation}
Perhaps one reason that there are comparatively few books on the Dirac
equation is the lack of an unambiguous quantum mechanical interpretation.
Dirac's electron theory seems to remain a theory with no clearly defined
range of validity, with peculiarities at its limits which are not completely
understood. Indeed, it is not clear whether one should interpret the Dirac
equation as a quantum mechanical evolution equation, like the Schr\"{o}%
dinger equation for a single particle. The main difficulty with a quantum
mechanical one-particle interpretation is the occurrence of states with
negative (kinetic) energy. Interaction may cause transitions to negative
energy states, so that there is no hope for a stability of matter within
that framework. In view of these difficulties R. Jost stated ``The
unquantized Dirac field has therefore no useful physical interpretation''.
Despite this verdict we are going to approach these questions in a pragmatic
way. A tentative quantum mechanical interpretation will serve as a guiding
principle for the mathematical development of the theory. It will turn out
that the negative energies anticipate the occurrence of antiparticles, but
for the simultaneous description of particles and antiparticles one has to
extend the formalism of quantum mechanics. Hence the Dirac theory may be
considered a step on the way to understanding quantum field theory.\cite%
{Thaller}
\end{quotation}

To speak of the Dirac equation is to speak mainly about relativistic quantum
mechanics (RQM). But, for this purpose, a replacement of Dirac's
mathematical formalism already exists. It is K\"{a}hler calculus. Written in
German, it has been under the radar except briefly a few decades ago for
highly specialized topics. And it has been overlooked that his superior
version of RQM is a virtual concomitant of his calculus; one only needs to
let the mathematics speak.

The most outstanding feature of the K\"{a}hler calculus (KC) is being the
refined, formally adequate representation of the exterior calculus cum
coderivative. This differentiation involves Hodge duality, operation which
does not belong to exterior algebra but to Clifford algebra. One should do
the calculus of differential forms in Clifford rather than exterior context.

The exterior calculus was born in 1899, almost in passing in an E. Cartan
paper \cite{R4}, in the area of differential equations known as Pfaff
systems. But it did not get traction for many decades. By his own admission
Chern's best paper\ is from 1944 \cite{ChernGB}. In it, he almost
apologetically, justifies his use of ``the theory of exterior differential
forms, instead of the ordinary tensor analysis ... '' as a matter of
convenience.

In 1960-1962, K\"{a}hler produced his calculus (KC) \cite{R46}, \cite{R47}
and \cite{R48}. In the last decade of his long life, he returned to this
topic \cite{K92}, not to the topic of K\"{a}hler metrics and K\"{a}hler
manifolds \cite{K33}, for which he is best known. A similar comment applies
to his work on Cartan-K\"{a}hler theory of exterior systems\cite{R45}. One
should perhaps take this as a statement of what he considered to be the most
important work of his life, or perhaps the one to which one should pay
greatest attention. The aim of this paper is to show that, even though the
direct and indirect applications of the exterior calculus are many and
important, they still pale in comparison with the fact that, in K\"{a}hler
form, it yields the magnificent version of RQM of which we speak in the
second half of this paper [For indirect applications, see for instance those
of K\"{a}hler manifolds, as described by Bourguignon \cite{Bourg}, and
numerous papers on global differential geometry by Chern].

In section 2, we shall give a brief description of the essence of the KC. In
Section 3, we give the proof of the equivalence of the coderivative and K%
\"{a}hler's interior derivative. In the interest of brevity, we assume that
readers know some exterior calculus and some Clifford algebra. It will not
escape their attention why the said proof reaches so far; any differential
form can be viewed as a member of both exterior and Clifford algebra.

Section 4 is devoted to enumerating several of the great features of the RQM
that emerges as a virtual concomitant of the KC. The exterior calculus in K%
\"{a}hler's version can thus appropriate itself of a RQM without the
difficulties of the Dirac theory. The future will be looking down on present
mathematicians that speak the language of Gauss, Grassmann, Hilbert, K\"{a}%
hler himself, F. Klein, Riemann and Weyl (to name just a few in a
constellation of mathematical luminaries over one and a half centuries) for
failing to understand the evolution of a mathematical line of development
that started with a Leibnizian prescience, namely his characteristica
geometrica \cite{OnLeibniz}.

\section{A Cartan approach to the KC}

K\"{a}hler introduced in ad hoc manner a concept of covariant derivative of
differential forms of tensor valuedness. For present purposes, we specialize
that formula to scalar-valuedness. It then reads
\begin{equation}
d_{h}a_{l_{1}\ldots l_{p}}=\frac{\partial }{\partial x^{h}}a_{l_{1}\ldots
l_{p}}-\Gamma _{hl_{1}}^{r}a_{rl_{2}\ldots l_{p}}-\ldots -\Gamma
_{hl_{p}}^{r}a_{l_{1}l_{2}\ldots l_{p-1}r},
\end{equation}%
where $a_{l_{1}\ldots l_{p}}$ stands for $a_{l_{1}\ldots
l_{p}}dx^{l_{1}}\wedge dx^{l_{2}}\wedge \ldots \wedge dx^{l_{p}}$. This
author is not responsible for this unfortunate approach and choice of
connection.

The KC of scalar-valued differential forms only requires a metric structure.
The affine (Euclidean, Lorentzian, etc.) structure is irrelevant. In 1922,
just before his paper on theory of affine connections, Cartan derived the
differential invariants that characterize a differentiable manifold endowed
with a metric.

A summary of his argument follows. He decomposes symmetric, quadratic
differential forms as sums of squares
\begin{equation}
ds^{2}=\sum^{n}\epsilon _{i}(\omega ^{i})^{2},\qquad (\epsilon _{i}=\pm 1).
\end{equation}%
The $\omega _{i}$'s are linear in the $dx^{i}$, but depend not only on the $%
x^{j}$ but also on $n(n-1)/2$ parameters $u$. Exterior differentials of the $%
\omega ^{i}$'s can be written as
\begin{equation}
d\omega ^{i}=\omega ^{j}\wedge \omega _{j}^{i},
\end{equation}%
where the $\omega _{j}^{i}$ are linear in the $dx$ and the $du$. We writes
down
\begin{equation}
(\omega ^{i})^{\delta }=0
\end{equation}%
to indicate that there are no $du^{\prime }$s in (2).

Cartan develops the consequences of (4) and finds
\begin{equation}
\omega _{ij}+\omega _{ji}=0.
\end{equation}%
The system of equations (3) and (5) uniquely defines the $\omega _{i}^{j}$
in the ``bundle manifold'', i.e. of the $(x,u)$, and on sections of this
bundle (manifold of the $x$'s). This system is familiar from the theory of
Euclidean connections, also improperly known as metric compatible affine
connections, which cease to be affine by virtue of the restriction on the
underlying group. But no connection has been used here. We shall write this
system as
\begin{equation}
d\omega ^{i}=\omega ^{j}\wedge \alpha _{j}^{i},\qquad \alpha _{ij}+\alpha
_{ji}=0,
\end{equation}%
reserving $\omega _{i}^{j}$ for the actual connection of a manifold, which
need not be the Levi-Civita connection. K\"{a}hler's derivative $\partial $
is conceived as
\begin{equation}
\partial u=\omega ^{\mu }\vee d_{\mu }u=du+\delta u
\end{equation}%
where
\begin{equation}
du=\omega ^{\mu }\wedge d_{\mu }u,\qquad \delta u=\omega ^{\mu }\cdot d_{\mu
}u
\end{equation}%
in the K\"{a}hler algebra, i.e. the Clifford algebra determined by
\begin{equation}
\omega ^{\mu }\vee \omega ^{\nu }+\omega ^{\nu }\vee \omega ^{\mu }=2g^{\mu
\nu },
\end{equation}%
$g^{\mu \nu }$ being the inverse matrix of the $g_{\lambda \rho }$, in turn
defined by $ds^{2}=g_{\mu \nu }\omega ^{\mu }\omega ^{\nu }.$

It is well known from Clifford algebra that
\begin{equation}
\omega^{\mu}\wedge \omega^{\nu} = \frac{1}{2}(\omega^{\mu}\vee \omega^{\nu}
- \omega^{\nu}\vee \omega^{\mu}),
\end{equation}
and
\begin{equation}
\omega^{\mu}\cdot \omega^{\nu} = \frac{1}{2}(\omega^{\mu} \vee \omega^{\nu}
+ \omega^{\nu}\vee \omega^{\mu}).
\end{equation}
Here we are not requiring that the $\omega^{\mu}$'s be orthonormal. We are
starting to use Greek indices as we shall reserve Latin indices for 3-space.

The Leibniz distributive rule without alternating signs is assumed. We shall
then only need $d_{\mu }\omega ^{\nu }$ in order to have $d_{\mu }u$ for any
scalar-valued differential form $u$. On sections of the bundle (i.e. the
manifold of the $x_{s}^{\prime }$ and $u^{\prime }s^{\prime \prime }$, the $%
\alpha _{\mu }^{\nu }$'s are not independent of the $\omega ^{\mu }$'s. We
write $\alpha _{\mu }^{\nu }=\Gamma _{\mu \;\lambda }^{\nu }\omega ^{\lambda
}$. Hence
\begin{equation}
d\omega ^{\mu }=\omega ^{\nu }\wedge \alpha _{\nu }^{\mu }=\omega ^{\nu
}\wedge \Gamma _{\nu \;\lambda }^{\mu }\omega ^{\lambda }=\omega ^{\lambda
}\wedge (-\Gamma _{\nu \;\lambda }^{\mu }\omega ^{\mu }).
\end{equation}%
For $u=\omega ^{\mu }$, the first of equations (8) becomes
\begin{equation}
d\omega ^{\mu }=\omega ^{\nu }\wedge d_{\nu }\omega ^{\mu }.
\end{equation}%
Comparison of (12) and (13) allows us to obtain two canonical $d_{\lambda }$
for $\omega ^{\mu }$, namely
\begin{equation}
d_{\nu }\omega ^{\mu }=\alpha _{\nu }^{\mu },\qquad d_{\lambda }\omega ^{\mu
}=-\Gamma _{\nu \;\lambda }^{\;\mu }\omega ^{\nu }.
\end{equation}%
A change of indices without consequence allows us to rewrite the second of
Eqs. (14) as:
\begin{equation}
d_{\nu }\omega ^{\mu }=-\Gamma _{\lambda \;\nu }^{\;\mu }\omega ^{\lambda }.
\end{equation}%
We have here two different covariant derivatives, namely the first of (14)
and (15). In terms of coordinate bases, $\Gamma _{\lambda \;\nu }^{\;\mu
}=\Gamma _{\nu \;\lambda }^{\mu }$. Then
\begin{equation}
d_{\nu }dx^{\mu }=-\Gamma _{\lambda \;\nu }^{\mu }dx^{\lambda }=-\Gamma
_{\nu \;\lambda }^{\mu }dx^{\lambda }=-\alpha _{\nu }^{\mu },
\end{equation}%
which explicitly shows that the two covariant derivatives are the opposite
of each other, in case this was not obvious from the moment that they were
introduced. We advance that, for notational compatibility with differential
geometry, we choose the option (15). We shall not enter the details but
point out at the fact that the standard divergence of a vector field is the
same as the interior derivative of a differential 1-form when one use (15).

\section{Identification of the covariant derivatives of the Cartan's and K\"%
{a}hler's approaches}

K\"{a}hler proceeded to simplify Eq. (1), by using that
\begin{equation}
\omega _{i}^{k}=\Gamma _{i\;j}^{\;k}dx^{j}
\end{equation}%
and obtained
\begin{equation}
d_{h}u=\frac{\partial u}{\partial x^{h}}-\omega _{h}^{r}\wedge e_{r}u,
\end{equation}%
where $\omega _{h}^{r}$ is our $\alpha _{h}^{r}$, where $e_{r}u$ is $%
=(dx)_{r}u$, and where $(dx)_{r}$ is defined by $g_{sr}dx^{r}$ (coordinates $%
dx_{r}$ do not exist except for rectilinear systems). In K\"{a}hler's
treatment, formula (18) is as ad hoc as the formula (1) form which he
obtained it. The appropriate process to get to (1), if that is what we want,
is to continue the process of the previous section and thus obtain (18), as
we are about to do.

Let $\omega ^{R}$ be the differentiable $r$-form $\omega ^{1}\wedge \omega
^{2}\wedge \ldots \wedge \omega ^{r}$. Let $s$ be $\leq r$ and let $\omega
_{(s)}$ be $(-1)^{s-1}\omega ^{1}\wedge \ldots \wedge \check{\omega}%
^{s}\wedge \ldots \wedge \omega ^{r}$, where $\check{\omega}^{s}$ means that
we have removed the factor $\omega ^{s}$ from $\omega ^{R}$, after making it
the first factor in the product. Clearly
\begin{equation}
\omega ^{R}=\omega ^{s}\wedge \omega _{(s)}\qquad (\text{no\ sum}).
\end{equation}%
Hence
\begin{equation}
\omega _{s}\cdot \omega ^{R}=\omega _{s}\cdot \lbrack \omega ^{s}\wedge
\omega _{(s)}]=(\omega _{s}\cdot \omega ^{s})\omega _{(s)}+0=\omega _{(s)},
\end{equation}%
if by $\omega _{\mu }$ we mean the elements of the basis reciprocal of the $%
\omega ^{\nu }$, i.e. $\omega _{\mu }\cdot \omega ^{\nu }=\delta _{\mu
}^{\nu }$. We thus have
\begin{equation}
d_{\mu }\omega ^{R}=\sum_{s=1}^{r}d_{\mu }\omega ^{s}\wedge \omega
_{(s)}=-\sum_{s=1}^{r}\Gamma _{\lambda \;\mu }^{\text{ }s}\omega ^{\lambda
}\wedge \omega _{(s)}=-\sum_{\sigma =1}^{n}\Gamma _{\lambda \;\mu }^{\text{ }%
\sigma }\omega ^{\lambda }\wedge \omega _{(\Sigma )},
\end{equation}%
where $\omega _{(\Sigma )}$ is $\omega _{(s)}$ for $\sigma \leq r$ and zero
for $\sigma >r.$ From (20) and (21), we get
\begin{equation}
d_{\mu }\omega ^{R}=-\sum_{\sigma =1}^{n}\Gamma _{\lambda \;\mu }^{\sigma
}\omega ^{\lambda }\wedge (\omega _{\sigma }\cdot \omega ^{R}).
\end{equation}%
Formula (22) applies for arbitrary bases of differential 1-forms. If the $%
\omega ^{\mu }$'s are $dx^{\mu }$'s, the $\Gamma _{\lambda \;\mu }^{\text{ }%
\sigma }$ are the Christoffel symbols. They satisfy $\Gamma _{\lambda \;\mu
}^{\text{ }\sigma }=\Gamma _{\mu \;\lambda }^{\text{ }\sigma }$. Thus
\begin{equation}
d_{\mu }(dx^{1}\wedge \ldots \wedge dx^{r})=-\alpha _{\mu }^{\sigma }\wedge
e_{\sigma }dx^{1}\wedge \ldots \wedge dx^{r}.
\end{equation}

We require $d_{\lambda }u$ to satisfy the Leibniz rule. Equation (23) and
the distributive property of $d_{\lambda }$ with respect to addition
together yield (18) (in coordinate bases!).

\section{The K\"{a}hler derivative as sum of exterior derivative and
coderivative}

It is implicit in K\"{a}hler's work that his derivative is the sum of the
exterior derivative and the coderivative. But the steps in the argument are
scattered over several sections containing a large amount of relevant
material. The proof is thus buried in his work, so much of it that one might
overlook the point that we are making in this paper.

Let $\eta $ be the linear operator that acting on differential r-forms, $%
u^{R}$, yields%
\begin{equation}
\eta u^{R}=(-1)^{r}u^{R}.
\end{equation}
Let $a$ a differential 1-form and let $A$ be an arbitrary element of the
Clifford algebra. Recall the well known relation
\begin{equation}
aA=a\wedge A+a\cdot A,
\end{equation}%
where
\begin{equation}
a\wedge A=\frac{1}{2}[aA+(\eta A)a]
\end{equation}%
and
\begin{equation}
a\cdot A=\frac{1}{2}[aA-(\eta A)a].
\end{equation}%
From (27) with $a=dx^{\mu }$ and $A=v$, we get
\begin{equation}
dx^{\mu }\vee v=(\eta v)\vee dx^{\mu }+2dx^{\mu }\cdot v,
\end{equation}%
to be used further below.

A differential form, $c$, is said to be constant if $d_{\mu }c=0.$ By virtue
of the Leibniz rule, we have
\begin{equation}
d_{\mu }(u\vee c)=(d_{\mu }u)\vee c.
\end{equation}%
We often use redundant notation (like using the parenthesis in this case)
for greater clarity. We have
\begin{equation}
\partial u=du+\delta u
\end{equation}%
where
\begin{equation}
du\equiv dx^{\mu }\wedge d_{\mu }u,\qquad \delta u\equiv dx^{\mu }\cdot
d_{\mu }u.
\end{equation}

We now proceed with the proof. Let $z$ be the unit differential n-form,
\begin{equation}
z=dx^{1}\wedge dx^{2}\wedge dx^{3}\wedge idx^{u}.
\end{equation}%
Recall that
\begin{equation}
\ast u=u\vee z,\quad \ast ^{-1}u=(-1)^{\binom{n}{2}}u\vee z,
\end{equation}%
and
\begin{equation}
\ast ^{-1}d\ast u=(-1)^{\binom{n}{2}}d(u\vee z)\vee z.
\end{equation}%
We use that $z$ is a constant differential and that
\begin{equation}
dx^{\mu }\cdot (u\vee v)=(dx^{\mu }\cdot u)\vee v+\eta u\vee (dx^{\mu }\cdot
v).
\end{equation}%
We replace $u$ and $v$ with $d_{\mu }u$ and $z:$%
\begin{align}
d(u\vee z)& =dx^{\mu }\wedge (d_{\mu }u\vee z)=dx^{\mu }\vee (d_{\mu }u\vee
z)-dx^{\mu }\cdot (d_{\mu }u\vee z)  \notag \\
& =\partial u\vee z-\delta u\vee z-(\eta d_{\mu }u)dx^{\mu }\vee z.
\end{align}%
Taking into account (28), the Clifford product of the last term of (36) by $%
z $ on the right becomes
\begin{align}
-(\eta d_{\mu }u)\vee dx^{\mu }\vee z& =-dx^{\mu }\vee d_{\mu }z\vee
z+2(dx^{\mu }\cdot d_{\mu }u)\vee z  \notag \\
& =-\partial u\vee z+2\delta u\vee z.
\end{align}%
Hence
\begin{equation}
d(u\vee z)=\partial u\vee z-\partial u\vee z-\partial u\vee z+2\delta u\vee
z=\partial u\vee z
\end{equation}%
and, therefore, (34) becomes
\begin{equation}
\ast ^{-1}d\ast u=(-1)^{\binom{n}{2}}(\delta u\vee z)z=\delta u.
\end{equation}%
We further have, from (30),
\begin{equation}
\partial u=du+\ast d\ast u,
\end{equation}%
as we wanted to prove.

\section{K\"{a}hler version of relativistic quantum mechanics}

K\"{a}hler's version of relativistic quantum mechanics (RQM) has gone
largely unnoticed. Worse yet is the fact that those who have cited his
papers on the subject ---a rare event in recent decades--- appear not to
have noticed its great advantages over Dirac's theory. We proceed the
document this.

The K\"{a}hler equation, to which he deferentially but improperly referred
as Dirac's equation, reads
\begin{equation}
\partial u=au,
\end{equation}%
where $a$ is some input scalar valued differential form and where $u$ is not
required to be a spinor but just a member of the Clifford algebra of
differential forms. Juxtaposition of symbols (in $au$ but not in $\partial u$%
) is here an alternative for the symbol $\vee .$ No valuedness other than
scalar is needed for present purposes.

K\"{a}hler used equation (41) to solve the hydrogen atom with little ``extra
effort''. To be sure, one does not reach the fine structure in a couple of
steps. But the effort involved lies in the development of rich theory of
structural nature, thus useful for other purposes. With the same equation
and also almost effortlessly, this author derived from (48) the Pauli-Dirac
equation and, in one more page, the Foldy-Wouthuysen Hamiltonian.

In Dirac's theory, one usually restricts oneself to electromagnetic
coupling. As pointed out by Thaller, the range of validity of that equation
is not clear. Whereas K\"{a}hler considered $\partial u=0$ to be the
equivalent of a Dirac equation, others would not think so. This equation
defines (strict)\ harmonicity, a topic that is not associated with Dirac,
except perhaps in some recondite publication. This is to be contrasted wiith
the fact that K\"{a}hler gave the title ``Integrals of the Dirac equation $%
\partial u=0$ in three dimensional Euclidean space'' to one of the sections
of this 1962 paper. In passing, we shall give here some inklings about
unusual applications, thus extended range of validity, of his equation.

The fact that the KC allows us to obtain results in a new, simpler way does
not have per se much more than anecdotal evidence. But there is the
transcendental feature of (41) that $u$ need to be a spinor, i.e. not belong
to an ideal of that algebra. So, it is not an equation for just particles of
spin $1/2$. It also applies to fields that have lost connection with any
specific particle.

Crucial for the important subject of conservation laws is the concept of
what he called scalar products of different grades, denoted as $(u,v)_{i}$,
the grade being $n-i.$ For $i=0,$ he wrote simply $(u,v)$. The first Green
identity reads
\begin{equation}
d(u,v)_{1}=(u,\partial v)+(v,\partial u).
\end{equation}%
If solutions $u$ and $v$ of an equation satisfy that the right hand side of
(42) is zero, a conservation law follows. Let overbar denote complex
conjugate. K\"{a}hler showed that, if $-\eta \overline{a}=a,$ and if $u$ and
$v$ are solutions of the K\"{a}hler equation with input $a$, then
conservation laws
\begin{equation}
d(u,\eta \overline{v})_{1}=0
\end{equation}%
follow and, in particular,
\begin{equation}
d(u,\eta \overline{u})_{1}=0.
\end{equation}%
We do not give the definition of $(u,v)_{1}$, as it would be an unnecessary
distraction here. Suffice to mention our use of the notation $<u|v>$ and $%
(u|v)$ for $(u,\eta v)_{1}$, respectively in the K\"{a}hler algebras built
upon the modules spanned by $(dt,dx^{i})$ and $(dx^{i})$, $i=1,2,3$.
Coefficients nevertheless depend on $t$ in both cases. But $\partial
/\partial t$ and thus $dt$ will be absent in the second case

Let $C$ be any constant element of the spacetime algebra such that $C^{2}$
(i.e. $C\vee C$) equals 1. The $\frac{1}{2}(1\pm C)$ are two mutually
annulling constant idempotents, $I^{\pm }.$ By virtue of (29),
\begin{equation}
\partial (u\vee I)=dx^{\mu }\vee d_{\mu }(u\vee I)=dx^{\mu }\vee d_{\mu
}u\vee I\equiv \partial u\vee I.
\end{equation}%
Equations (41) and (45) together imply
\begin{equation}
\partial (u\vee I)=\partial u\vee I=au\vee I=a(u\vee I),
\end{equation}%
which shows that the $u\vee I^{\pm }$ are solutions of the same K\"{a}hler
equation as the $u$'s. Because their sum is unity, we have \
\begin{equation}
u=u\vee I^{+}+u\vee I^{-}
\end{equation}%
And because they mutually annul, multiplication of (47) respectively by $%
I^{+}$ and $I^{-}$ uniquely defines ${^{+}u}$ and ${^{-}u}$ in
\begin{equation}
u=^{+}\!u\vee I^{+}+^{-}\!u\vee I^{-}.
\end{equation}%
Since $(dt)^{2}=-1$, we have the constant idempotents
\begin{equation}
\epsilon ^{\pm }=\frac{1}{2}(1\pm idt).
\end{equation}%
We take exception to not exhibiting the unit imaginary because ''$i$'' is of
the essence here.

We next assume that the input $a$ of the K\"{a}hler equation satisfies $\eta
\overline{a}=a.$ This is the case in particular for electromagnetic
coupling, $-iE_{0}+e\omega $, where $E_{0}$ is rest mass and where $e=\mp
\left| e\right| $ is the charge of electron/positron. K\"{a}hler uses that
\begin{equation}
u=^{+}\!u\vee \epsilon ^{+}+^{-}\!u\vee \epsilon ^{-},
\end{equation}%
but not necessarily stationarity. After some calculations, he gets
\begin{align}
<u|u>& =\frac{1}{2}(^{+}u,^{+}\!\overline{u})+\frac{1}{2}(^{+}\!u|\eta ^{+}\!%
\overline{u})\wedge idt  \notag \\
& -\frac{1}{2}(^{-}u,-\overline{u})+\frac{1}{2}(^{-}u|\eta ^{-}\overline{u}%
)\wedge idt.
\end{align}%
It goes without saying that, if
\begin{equation}
d(u,\eta \overline{u})_{1}=d<u|u>=0,
\end{equation}%
a conservation law of the form
\begin{equation*}
d(j^{(1)}+j^{(2)})=0,
\end{equation*}%
follows. The $j^{(1)}$ and $j^{(2)}$ are spacetime currents that come with
corresponding densities, $\pm \frac{1}{2}(^{\pm }u,^{\pm }\!\overline{u}).$

The sign definiteness of the charge densities, which are of identical form
except for sign, speaks of the fact that, in K\"{a}hler's quantum mechanics,
the wave function is about amplitude of charge density, not of probability
density. Of course, the Copenhagen interpretation will still work in the
situations in which it is applied, but not because of being a basic tenet of
this quantum mechanics, where it is not a fundamental but a derived tenet.

Endowed with this interpretation of what the two terms in (50) are when $%
\eta \overline{a}$ equals $a,$ let us assume stationarity. K\"{a}hler then
writes $u$ as
\begin{equation}
u=p^{+}\!\vee T^{+}+p^{-}\vee T^{-},
\end{equation}%
where
\begin{equation}
T^{\pm }=e^{-itE/\hbar }\epsilon ^{\pm }.
\end{equation}%
Though K\"{a}hler could have considered solutions $p^{+}\vee T^{+}$ and $%
p^{-}\vee T^{-}$independently of each other, he chose not to do so. There is
no need for that as, the equation for $u$ immediately splits into the
equations
\begin{equation}
\partial p^{\pm }\pm \left( \frac{E}{\hbar }+\beta \right) \vee \eta p^{\pm
}-\alpha \vee p^{\pm }=0
\end{equation}%
after decomposing $a$ as $\alpha +\beta \vee idt.$ The two equations differ
only by the sign of the second term. Notice that we have two equations for
the same $E$, not%
\begin{equation}
\partial p^{\pm }+\left( \pm \frac{E}{\hbar }+\beta \right) \vee \eta p^{\pm
}-\alpha \vee p^{\pm }=0.  \tag{NOT!}
\end{equation}

We emphasize that (55) is for coupling more general than electromagnetic, in
which case they would be for positions and electrons respectively. Notice in
(57) that they are associated with the same sign of the energy. Notice
further that there is neither room no need to eliminate small components of
the wave function as if they belonged to an antiparticle contamination of
the wave function. As is well known, that is the case in Dirac's theory.

What we have said so far is part of the twelve pages on RQM in his paper %
\cite{R48}, part of which is devoted to the fine structure of the $H$ atom.
In a previous paper \cite{R47}, he had solved the same problem starting with
proper solutions for energy and momentum. For that purpose, he defined
idempotents $\tau ^{\pm }$:
\begin{equation}
\tau ^{\pm }=\frac{1}{2}(1\pm idxdy).
\end{equation}%
They add up to unity, kill each other and commute with $\epsilon ^{\ast }$.
This allows him to write
\begin{equation}
u=^{+}\!u^{+}\vee \tau ^{+}\vee \epsilon ^{+}+^{+}\!u^{-}\vee \tau ^{-}\vee
\epsilon ^{+}+^{-}\!u^{+}\vee \tau ^{+}\vee \epsilon ^{-}+^{-}u^{-}\vee \tau
^{-}\vee \epsilon ^{-},
\end{equation}%
the $^{\pm }u^{\ast }$ being all well defined in terms of $u$:
\begin{equation}
^{\pm }u^{\ast }=u\vee \tau ^{\ast }\vee \epsilon ^{\pm }.
\end{equation}%
One can associate $\tau ^{\pm }$ with spin/chirality in the same way as $%
\epsilon ^{\pm }$ is associated with energy/charge. K\"{a}hler then assumed
central fields, apparently so that the system of a cylindrically symmetric
particle in a field will not lose this symmetry by virtue of the
non-cylindrically symmetric field. He wrote the ansatz
\begin{equation}
u=e^{im\phi -iEt/\hbar }p\vee \tau ^{\pm }\vee \epsilon ^{\ast },
\end{equation}%
where $p$ depends on $(\rho ,z,d\rho ,dz)$, but not on $(t,\phi ,dt,d\phi )$%
, where $\phi $ is angular cylindrical coordinate and where $m$ is angular
momentum. In view of the combined (57) and (59) equations, it is clear that
both positrons and electrons fit together at the same time in the\ K\"{a}%
hler equation.

Another great result of K\"{a}hler version of RQM is his treatment of
angular momentum, which he approaches from a perspective of Lie
differentiation. But this differentiation is not what one would expect, as
vector fields and their flows do not enter his concept of Lie
differentiation (unless one defines vector fields as $\partial /\partial
x^{\mu }$ operators and their linear combinations, which neither Cartan not K%
\"{a}hler do). The major though not unique result of K\"{a}hler's treatment
of angular momentum, $\partial /\partial \phi $, is that both orbital and
spin angular momentum are unified ab initio, actually before being born, the
meaning of which we shall explain below.

K\"{a}hler refers to the operators (which cyclic $i,j,k$)
\begin{equation}
X_{i}=x^{j}\frac{\partial }{\partial x^{k}}-x^{k}\frac{\partial }{\partial
x^{j}},
\end{equation}%
as Lie operators (see formula 22.2 of \cite{R48}). All the three $X_{i}$ are
of the form $\partial /\partial \phi _{i}$, where $\phi _{i}$ is azimuthal
coordinate with respect to the three axes. With $w_{i}\equiv dx^{j}\wedge
dx^{k}$, the action of $X_{i}$ on $u$ is%
\begin{equation}
X_{i}u=x^{j}\frac{\partial u}{\partial x^{k}}-x^{k}\frac{\partial u}{%
\partial x^{j}}+\frac{1}{2}w_{i}\vee u-\frac{1}{2}u\vee w_{i}.
\end{equation}%
The argument leading to this equation from ``first principles'' is given in
section 33 of \cite{R46}.

We proceed to illustrate how the argument goes. Given operator $X$,
\begin{equation}
X=\alpha ^{i}(x^{1},\ldots ,x^{n})\frac{\partial }{\partial x^{i}}.
\end{equation}%
K\"{a}hler introduces the differential system
\begin{equation}
\frac{dx^{i}}{dy^{n}}=\alpha ^{i}(x^{1},\ldots ,x^{n}),
\end{equation}%
which is a familiar system from classical mechanics, where $y^{n}$ is time, $%
t$. The reason for using the symbol $y^{n}$ is that the $n-1$ independent
constants of the motion not additive to $y^{n}$ together with $y^{n}$
constitutes a new coordinate system, $y^{i}$, and thus
\begin{equation}
x^{i}=x^{i}(y^{1},\ldots ,y^{n}).
\end{equation}%
In terms of the $y$ coordinate system $X$ reduces to $\partial u/\partial
y^{n}$. K\"{a}hler then computes $\partial u/\partial y^{n}$ with $%
u=a_{R}dx^{R}$ and obtains
\begin{equation}
\frac{\partial u}{\partial y^{n}}=(Xa_{R})dx^{R}+d\alpha ^{i}\wedge e_{i}u.
\end{equation}%
He has resorted to coordinates $(y^{i})$ because the different terms on the
right of (62) correspond to different conditions. So application of the sum
is not equivalent to sum of the different partial derivatives. On the other
hand, $\partial u/\partial y^{n}$ is just one partial derivative, not a sum
of them. It is clear that we have
\begin{equation}
\frac{\partial u}{\partial y^{n}}=\alpha ^{i}\frac{\partial u}{\partial x^{i}%
}+d\alpha ^{i}\wedge e_{i}u
\end{equation}%
for arbitrary differential forms $u$.

K\"{a}hler refers to the right hand side of (66) as the Lie derivative of $u$%
. But, by his argument, it is simply the partial derivative, $\partial
u/\partial y^{n}.$ One does not need to explicitly compute $y^{n}.$ Equation
(66) directly gives as its form in term of the original coordinate system. A
simple example will help with another remark.

The third component of angular momentum is $x\frac{\partial }{\partial y}-y%
\frac{\partial }{\partial x}$. Two (of infinite) coordinate system
containing $\phi $ (i.e. the corresponding $y^{n}$) are the spherical and
cylindrical ones. $\phi $ is determined by its coordinate line, which is
independent of the other coordinates, $r$ and $\theta $ or $\rho $ and $z$.
Hence $\partial /\partial y^{n}$ is determined neither by a specific allowed
$(y^{i})$ system, nor by what coordinate system we choose to express $u$
(Cartesian coordinates in our example).

With respect to the unification of the two types of angular momentum,
orbital and intrinsic, consider the following. The terms in (66) are not the
same in the $x$ and $x^{\prime }$ coordinate systems (This is remedied by
the use of covariant derivatives). Only their sum is. One can add and
subtract the appropriate quantity to the respective terms, so that the sum
is unchanged and the two terms on the right are covariant. (66) then becomes
\begin{equation}
Xu=\alpha ^{l}d_{l}u+(d\alpha )^{l}\wedge e_{l}u.
\end{equation}%
K\"{a}hler uses the symbol $(d\alpha )^{i}$ to represent what may be viewed
as components as a vector of the differential form $d(\alpha ^{i}%
\boldsymbol{e})$ under the Levi-Civita connection (This remark is only for
identification purposes of what quantities are involved; the manifold may
not even be endowed with a connection, i.e. a rule to compare vectors in two
tangent spaces).

For $X_{l},$the first (second) term in (67) will yield the first two terms
(respectively third and fourth) terms in (64). They represent orbital and
intrinsic angular momenta. They are ``entangled'' in (66) before they became
individually meaningful and interpreted as angular momentum concepts in
(64). One could not have deeper unification than that.

Equation (67) involves only the exterior product. K\"{a}hler developed this
equation further when the metric does not depend on the coordinate $y^{n}$.
The Clifford product emerges in the expression for the Lie derivative. He
further specializes it to spacetime. That is how equation (64) results.

K\"{a}hler defines total angular momentum not as a vector or bivector
operator, but through
\begin{equation}
(K+1)u=\sum_{i=1}^{3}X_{i}u\vee w^{i}.
\end{equation}%
The reason for using $k+1$ on the left rather than just $M$ (i.e. $M=K+1$)
is that, as he proves
\begin{equation}
K(K+1)=-(X_{1}^{2}+X_{2}^{2}+X_{3}^{2}).
\end{equation}%
One should not worry about the minus sign. It would not appear if we had
multiply by $i$ in the definition of $X^{i}$. Notice that we could have
written the right hand side of (68) as
\begin{equation}
\frac{\partial u}{\partial \phi ^{i}}dx^{j}dx^{k}
\end{equation}%
where $\phi ^{i}$ pertains to the plane $(x^{j},x^{k})$. Hence total angular
momentum is the differential 2-form operator whose components are the $%
\partial /\partial \phi ^{i}$. This remark is helpful in understanding the
presence of the first exponential and the factor $\tau $ in (59).

We have devoted a lot of text to Lie differentiation and angular momentum.
It is the price we have to pay for illustrating the many subtleties
connected with these concepts. More enticing possibilities are intimated in
the next section.

\section{Loose ends and the mathematical owner of relativistic quantum
mechanics}

In papers \cite{R46}, \cite{R47} and \cite{R48}, K\"{a}hler gave not a name
to his underlying algebra. In an additional paper of 1964 \cite{K62b}, he
summarized results from the aforementioned papers and gave the name of
Clifford to the algebra he had been using, and again in 1992 \cite{K92}.

We have only focussed on the relation between the coderivative and the
interior derivative. We take it for granted that readers know how to relate
exterior products to Clifford products and vice versa. K\"{a}hler's formula
(9.1) of \cite{R48} expresses the Clifford product of two arbitrary elements
of the algebra in terms of exterior products. In particular, one can make
one of the two factors be the unity.

For the reverse process, i.e. to show that the exterior product can be
written in terms of Clifford products, suffice to show that such is the case
for the product of two monomials,
\begin{align}
u\wedge v& =(adx^{1}\wedge \ldots \wedge dx^{r})\wedge (bdx^{r+1}\wedge
\ldots \wedge dx^{s})  \notag \\
& =abdx^{1}\ldots \wedge dx^{r}\wedge dx^{r+1}\ldots \wedge dx^{s}.
\end{align}%
The proof then goes as follows. Let $Y^{i}\equiv dx^{i+1}\wedge \ldots
\wedge dx^{s}$. Then
\begin{equation}
dx^{1}\wedge Y^{1}=\frac{1}{2}[dx^{1}\vee Y^{1}+(\eta Y^{1})\wedge dx^{1}].
\end{equation}%
One can express $Y^{1}$ as
\begin{equation}
\frac{1}{2}[dx^{2}\vee Y^{2}+(\eta Y^{2})\vee dx^{2}]
\end{equation}%
and proceed similarly with $\eta Y^{2}$, and then $Y^{3}$, etc. This should
be enough to justify our claim.

There is a K\"{a}hler calculus more comprehensive than the one we have
discussed. It deals with tensor-valued differential forms. K\"{a}hler
considers the curvature as a tensor-valued differential 2-form. But
Euclidean curvatures are bivector-valued differential 2-forms, as Cartan
already stated \cite{R11}. Bivectors are not antisymmetric tensors since the
tensor product of antisymmetric tensors is not an antisymmetric tensor in
general. Curvatures are members of quotient algebras (not subalgebras) of
the general tensor algebra. We do not find tensor-valuedness to be
interesting at all.

There is, however, a very interesting point in K\"{a}hler's dealing with
tensor-valued differential forms. Their components have three series of
indices, two of which are for subscripts. One of these two are for covariant
tensors, whether antisymmetric or not. The other one is for differential
forms viewed as functions of r-surfaces. For Cartan and K\"{a}hler,
differential forms are not antisymmetric r-linear functions of vectors. The
differentiation of scalar-valued differential forms is determined by the
Christoffel symbols. The differentiation of linear functions of vectors is
determined by the connection, whose components will not be given by those
symbols when there is torsion.

K\"{a}hler's version of RQM is a concomitant of his calculus, which, in
turn, is nothing but the exterior calculus cum coderivative reformulated so
that the underlying algebra is manifestly Clifford algebra, rather than
exterior algebra that is complemented with non-exterior concepts like Hodge
duality and coderivative. Hence (the K\"{a}hler version of) RQM, which we
have just shown to be superior to Dirac's, may be said to be owned by the
calculus of scalar-valued differential forms. The Dirac calculus, a
magnificent achievement of the first third of the twentieth century, is
nowadays unnecessary and should be abandoned.

\section{Beyond the relation between the exterior calculus and relativistic
quantum mechanics}

At present, RQM has not a well defined boundary with quantum field theory.
The latter can be considered as an extension of the former, but one may
argue that is not natural nor canonical. There are topics which some authors
consider as pertaining to RQM and that other authors consider as pertaining
to its operator based extension. And there is also the S matrix theoretical
alternative to quantum field theory.

Because of its unparalleled contribution to quantum physics since 1948, let
us mention that Schwinger finds problems with quantum field theory that his
proposed source theory does not have \cite{Schwinger}. He also tells us how
sources imitate but supersede S matrix theory, ibid. Sources have much of
the flavor of differential forms, but they still are an ad hoc construction.
It is a concomitant of the main result of this paper that, given the
relation between the exterior calculus and RQM, one should look for an
extension of RQM in the extension of the exterior calculus.

The natural extension of the exterior calculus is differential geometry,
which Cartan and many differential geometers view as the exterior calculus
of vector-valued forms (the bivector valuedness comes in the wash when the
manifold is endowed with a metric). The natural extension of
scalar-valuedness then is vector-valuedness, and some algebra built upon the
module of vector fields. This takes place very simply through the
replacement of the unit imaginary with ``mirror elements in the tangent
algebra of differential forms in the idempotents. Thus $idxdy$ should be
replaced with $\mathbf{ij}dxdy.$ For further details, see a series of papers
posted in arXiv, where I have started the replacement of the unit imaginary
with elements of a tangent Clifford algebra (Type Jose G. Vargas on the
right hand corner of the arXiv's main page and ignore the entries with
multiple authorship), We now give an inkling of what to gain with such a
replacement.

The two differential forms $idt$ and $idxdy$ give rise to the eight
idempotents $\epsilon ^{\pm }$, $\tau ^{\pm }$ and $\epsilon ^{\pm }\tau
^{\ast }.$ With one more square root of one, we could build idempotents $%
\epsilon ^{\pm }$, $\tau ^{\pm }$, $\lambda ^{\pm }$, $\epsilon ^{\pm }\tau
^{\ast }$, $\epsilon ^{\pm }\lambda ^{\ast },$ $\tau ^{\pm }\lambda ^{\ast }$
and $\epsilon ^{\pm }\tau ^{\ast }\lambda ^{\ast \ast }.$ There are $2^{3}$
of them just of the type $\epsilon ^{\pm }\tau ^{\ast }\lambda ^{\ast \ast
}. $ What could the $\lambda ^{\pm }$ be? We skip considering $(1/2)(1\pm
idydz) $ since the difference with $\tau ^{\pm }$ is just a choice of
coordinates. Let us notice in passing that the $(1/2)(1\pm idydz)$ do not
commute with the $\tau ^{\pm }$, but $(1/2)(1\pm \mathbf{ij}dxdy)$ and $%
(1/2)(1\pm \mathbf{jk}dydz).$

Consider also idempotents of the type $\lambda _{i}^{\pm }=(1/2)(1\pm
\mathbf{a}_{i}dx^{i})$, with no sum over repeated indices. The three $%
\lambda _{i}^{\pm }$ commute among themselves and with some of the
idempotents previously considered. The issue arises of what the presence or
absence of commutativity implies.

Finally, let $\epsilon _{1}^{\pm }$, $\epsilon _{2}^{\pm }$, $\epsilon
_{3}^{\pm }$, ... $\epsilon _{r}^{\pm }$ be ``monary'', i. e not products of
other ones. We can perform all sorts of decomposition of the unity in terms
of them. Thus, for instance,%
\begin{equation}
1=\epsilon _{1}^{+}+\epsilon _{2}^{-}=\epsilon _{1}^{+}+\epsilon
_{2}^{-}\epsilon _{5}^{+}+\epsilon _{2}^{-}\epsilon _{5}^{-}=\epsilon
_{1}^{+}+(\epsilon _{2}^{-}\epsilon _{5}^{+}\epsilon _{6}^{+}++\epsilon
_{2}^{-}\epsilon _{5}^{-}\epsilon _{6}^{-})+\epsilon _{2}^{-}\epsilon
_{5}^{-}=...
\end{equation}%
where we have introduced parenthesis for greater clarity. If particles are
associated with idempotents, decompositions of this type are a golden rule
for creating all sorts of plausible particle reactions. We could refer to
this mechanism of particle stoichiometry.

That is just an example of what a KC of Clifford-valued differential forms
(i.e. members of the tensor product of two Clifford algebras) could do! If
you have contrasting ideas, it all will be for the benefit of mathematics.

\end{document}